This is the accepted (postprint) version of the following article: Besserer-Offroy É, *et al.* (2017). Eur J Pharmacol. doi: 10.1016/j.ejphar.2017.03.046, which has been accepted and published in its final form at http://www.sciencedirect.com/science/article/pii/S0014299917302157# The signaling signature of the neurotensin type 1 receptor with endogenous ligands

Élie Besserer-Offroy Offroy[ORCID ID], Rebecca L. Brouillette, Sandrine Lavenus, Ulrike Froehlich, Andrea Brumwell, Alexandre Murza, Jean-Michel Longpré, Éric Marsault[ORCID ID], Michel Grandbois, Philippe Sarret*, and Richard Leduc*[ORCID ID]

Department of Pharmacology-Physiology, Faculty of Medicine and Health Sciences, Institut de Pharmacologie de Sherbrooke, Université de Sherbrooke, Sherbrooke, Québec, Canada J1H 5N4

*e-mail addresses of the authors:*

Elie.Besserer-Offroy@USherbrooke.ca (ÉBO); Rebecca.Brouillette@USherbrooke.ca (RLB); Sandrine.Lavenus@USherbrooke.ca (SL); Ulrike.Froehlich@USherbrooke.ca (UF); Andrea.Brumwell@USherbrooke.ca (AB); Alexandre.Murza@USherbrooke.ca (AM); Jean-Michel.Longpre@USherbrooke.ca (JML); Eric.Marsault@USherbrooke.ca (ÉM); Michel.Grandbois@USherbrooke.ca (MG); *Philippe.Sarret@USherbrooke.ca (PS); *Richard.Leduc@USherbrooke.ca (RL)

**\* Co-corresponding authors:**

Philippe Sarret, PhD and Richard Leduc, PhD

Phone: +1 (819) 821-8000 ext. 75413

Fax: +1 (819) 564-5400© 2017. This manuscript version is made available under the CC-BY-NC-ND 4.0 license http://creativecommons.org/licenses/by-nc-nd/4.0/ 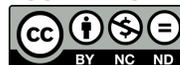

**Abstract**

The human neurotensin 1 receptor (hNTS1) is a G protein-coupled receptor involved in many physiological functions, including analgesia, hypothermia, and hypotension. To gain a better understanding of which signaling pathways or combination of pathways are linked to NTS1 activation and function, we investigated the ability of activated hNTS1, which was stably expressed by CHO-K1 cells, to directly engage G proteins, activate second messenger cascades and recruit β-arrestins. Using BRET-based biosensors, we found that neurotensin (NT), NT(8-13) and neuromedin N (NN) activated the $G\alpha_q$-, $G\alpha_{i1}$-, $G\alpha_{oA}$-, and $G\alpha_{13}$-protein signaling pathways as well as the recruitment of β-arrestins 1 and 2. Using pharmacological inhibitors, we further demonstrated that all three ligands stimulated the production of inositol phosphate and modulation of cAMP accumulation along with ERK1/2 activation. Interestingly, despite the functional coupling to $G\alpha_{i1}$ and $G\alpha_{oA}$, NT was found to produce higher levels of cAMP in the presence of pertussis toxin, supporting that hNTS1 activation leads to cAMP accumulation in a $G\alpha_s$-dependent manner. Additionally, we demonstrated that the full activation of ERK1/2 required signaling through both a PTX-sensitive $G_{i/o}$-c-Src signaling pathway and PLCβ-DAG-PKC-Raf-1-dependent pathway downstream of $G_q$. Finally, the whole-cell integrated signatures monitored by the cell-based surface plasmon resonance and changes in the electrical impedance of a confluent cell monolayer led to identical phenotypic responses between the three ligands. The characterization of the hNTS1-mediated cellular signaling network will be helpful to accelerate the validation of potential NTS1 biased ligands with an improved therapeutic/adverse effect profile.

Neurotensin; Neuromedin N

**Chemical compounds studied in this article:**

Neurotensin (1-13) (PubChem CID: 25077406); Neurotensin (8-13) (PubChem CID: 5311318);

Neuromedin N (PubChem CID: 9940301); SR48692 (PubChem CID: 119192)

**1. Introduction**

G protein-coupled receptors (GPCRs) constitute the largest superfamily of cell-surface proteins involved in signal transduction (Bockaert, 1999). A large and diverse group of ligands activates GPCRs, which transduce intracellular signals by coupling to G proteins or other proteins such as arrestins (Marinissen and Gutkind, 2001). Today, GPCRs are the targets of up to 40% of the total drug market and hence continue to be of great interest for the development of therapeutic agents.

Given that a GPCR can couple to different signaling pathways and that the latter can be selectively engaged or blocked led to the emergence of a novel paradigm called bias signaling (Kenakin, 2013). However, to fully exploit bias signaling, a clear profile of the signaling pathways activated by a GPCR is crucial. Here, we focused on deciphering the signaling signature of human NTS1 following stimulation by endogenous ligands. The neurotensin receptor family is composed of three subtypes, NTS1, NTS2, and NTS3 (Vincent et al., 1999). Among these subtypes, NTS1 (Tanaka et al., 1990; Vita et al., 1993) and NTS2 (Chalon et al., 1996; Mazella et al., 1996; Vita et al., 1998) belong to the rhodopsin-like family of GPCRs, whereas NTS3 belongs to the sortilin receptor family (Mazella and Vincent, 2006; Mazella et al., 1998). Since its discovery, NTS1 has been detected in peripheral tissues such as the vascular endothelium and gastrointestinal tract (Azriel and Burcher, 2001; Coppola et al., 2008) as well as in the central nervous system (CNS) (Roussy et al., 2008), suggesting that it is involved in a broad variety of physiological functions (Feng et al., 2015; Osadchii, 2015). Indeed, NTS1 mediates blood pressure lowering (Rioux et al., 1982), ileum contraction or relaxation (Carraway and Mitra, 1994), analgesia (Roussy et al., 2008), and hypothermia (Feifel et al., 2010).

NTS1 is endogenously activated by neurotensin (NT) as well as by neuromedin N (NN), both of which are derived from the same pro-NT/NN precursor (Rostène and Alexander, 1997; Rovere et al., 1996). Structure-activity studies of NT have revealed that the minimal bioactive fragment corresponds to the C-terminal hexapeptide of NT, NT(8-13) (St-Pierre et al., 1981).

To date, depending on the cell type, NTS1 has been linked, through $G\alpha_{i/o}$, $G\alpha_{q}$, and $G\alpha_{s}$ coupling, to a variety of intracellular signaling cascades including cyclic AMP (cAMP), inositol phosphate (IP), and arachidonic acid accumulation as well as the activation/inhibition of mitogen-activated protein kinases (ERK1/2 and JNK) and serine/threonine protein kinase Akt (Müller et al., 2011). Interestingly, NTS1-G protein interactions involve different receptor domains; interaction with $G\alpha_{q/11}$ requires the intact third intracellular loop of the receptor, and coupling to $G\alpha_{s}$ and $G\alpha_{i/o}$ involves the receptor's C-terminal domain (Gailly et al., 2000; Grisshammer and Hermans, 2001; Najimi et al., 2002; Skrzydelski et al., 2003). NT binding to NTS1 also induces the internalization of the receptor–ligand complex via a clathrin-dependent endocytic process requiring the recruitment of dynamin, intersectin, and β-arrestins to the GRK2/GRK5 phosphorylated receptor (Inagaki et al., 2015; Oakley et al., 2001; 2000; Savdie et al., 2006; Vandenbulcke et al., 2000; Zhang, 1999). Like many GPCRs, NTS1 internalization has been shown to lead to G protein-independent signaling (Souazé et al., 1997; Toy-Miou-Leong et al., 2004). Although the coupling to diverse G proteins has been demonstrated through the modulation of second messenger cascades, there is still no direct evidence of the interaction between NTS1 and G proteins in a unique cell type.

In the present study, we used bioluminescence resonance energy transfer (BRET) biosensors and label-free whole-cell assays to study the ability of NT, NT(8-13) and NN to activate signaling pathways following their binding to hNTS1 expressed in CHO-K1 cells.

## 2. Materials and Methods

### *2.1 Materials*

Neurotensin 1-13 (NT), IBMX (3-isobutyl-1-methylxantine), SR48692, and forskolin were obtained from R&D Systems, Inc. (Minneapolis, MN, USA). NT(8-13) was synthesized at the peptide synthesis facility of the Université de Sherbrooke. NN was purchased from GenScript USA Inc. (Piscataway, NJ, USA). Coelenterazine 400A (DeepBlueC) was purchased from Gold Biotechnology Inc. (St. Louis, MO, USA). The peptide sequences of NT, NT(8-13) and NN are shown in **Table 1**. Ham's F12, HEPES (4-(2-hydroxyethyl)-1-piperazineethanesulfonic acid), Penicillin-streptomycin-glutamine and fetal bovine serum (FBS) were obtained from Wisent (St. Bruno, QC, Canada), Opti-MEM was acquired from Invitrogen (Burlington, ON, Canada). Lance Ultra cAMP and pERK1/2 assay kits were purchased from Perkin Elmer (Montréal, QC, Canada). A Cisbio IP1 assay kit was purchased through Cedarlane (Burlington, ON, Canada). All inhibitors used in this study were from Santa-Cruz Biotechnologies (Dallas, TX, USA), with the exception of UBOQIC, which was purchased from Bonn University; Pertussis Toxin (PTX) which was purchased from List Biological Laboratories (Campbell, CA, USA); Dynasore, which was obtained from Tocris (Minneapolis, MN, USA); and Y27632, which was obtained from Cell Signaling Technology (Danvers, MA, USA).

### *2.2 Plasmids and constructs*

The cDNAs encoding the human neurotensin receptor type 1 and the $G_{\beta 1}$ subunit were obtained from the Missouri S&T cDNA Resource Center (Rolla, MO, USA). The fusion vector pIREShygro3-GFP10 and RlucII-β-arrestin 1 or 2 plasmids were kindly provided by Dr. Michel

Bouvier (Dept. of Biochemistry and IRIC, Université de Montréal, Montréal, QC, Canada). Using an InFusion advantage PCR cloning kit (Clontech Laboratories, Mountain View, CA, USA), the hNTS1 construct was inserted into the pIREShygro3-GFP10 vector as previously described (Demeule et al., 2014). The plasmids encoding $G\alpha_q$-RlucII (Breton et al., 2010), $G\alpha_{oA}$-RlucII (Richard-Lalonde et al., 2013), $G\alpha_{13}$-RlucII (Demeule et al., 2014), $G\alpha_{i1}$-RlucII, GFP10-$G\gamma_1$, and GFP10-$G\gamma_2$ (Galés et al., 2006) were kindly provided by Dr. Michel Bouvier. All constructs were verified by DNA sequencing.

## 2.3 Cell culture and transfections

CHO-K1 cells stably expressing hNTS1 (CHO-hNTS1 cells) were purchased from Perkin Elmer (Montréal, QC, Canada) and cultured in Ham's F12 containing 20 mM HEPES, 10% FBS, 0.4 mg/ml G418, and penicillin (100 U/ml)-streptomycin (100 µg/ml)-glutamine (2 mM) under 5% $CO_2$ at 37°C in a humidified atmosphere. The CHO-K1 cells were cultured in the same conditions as above but without G418. For the transient expression of recombinant proteins, T75 flasks were seeded with $3 \times 10^6$ cells, and 24 h later, the cells were transfected using PEI (Ehrhardt et al., 2006).

## 2.4 BRET$^2$ assay

To monitor direct G protein activation, we used the following biosensor couples: $G\alpha_q$-RlucII, GFP10-$G\gamma_1$, and $G\beta_1$ (Breton et al., 2010); $G\alpha_{oA}$-RlucII, GFP10-$G\gamma_1$, and $G\beta_1$ (Richard-Lalonde et al., 2013); $G\alpha_{13}$-RlucII, GFP10-$G\gamma_1$, and $G\beta_1$ ; or $G\alpha_{i1}$-RlucII, GFP10-$G\gamma_2$, and $G\beta_1$ (Galés et al., 2006). G protein biosensors were transfected into CHO-hNTS1 cells. At 24 h post-transfection, the cells were detached with trypsin-EDTA and plated (50,000 cells/well) in white opaque 96-well plates (BD Falcon, Corning, NY, USA). At 48 h post-transfection, the cells were washed once with PBS, and 90 µL of HBSS containing 20 mM HEPES was then added. Ligands were added at increasing concentrations for 20 min followed by coelenterazine-400A (5 µM). The BRET$^2$ measurements were collected in the 400 to 450 nm window (RlucII) and in the 500 to 550 nm window (GFP10) using the BRET$^2$ filter set on a GENios Pro plate reader (Tecan, Durham, NC, USA). The BRET$^2$ ratio was determined as the light emitted by the acceptor GFP10 over the light emitted by the donor RlucII. The monitoring of β-arrestin recruitment was done by the transient transfection of CHO-K1 cells with plasmids containing cDNAs encoding hNTS1-GFP10 and RlucII-β-arrestin 1 or 2. The same protocol as the one used for G protein activation was then used.

### 2.5 IP-One and Lance Ultra cAMP assays

The IP-One assay was performed according to the manufacturer's recommendations. Briefly, 15,000 CHO-hNTS1 cells per well (384-well shallow well plate) were treated with increasing concentrations of NT, NT (8-13), or NN for 30 min. IP1-d2 and anti-IP1-Cryptate were added for at least 1 h. The plates were read on a GENios Pro plate reader with HTRF filters (excitation at 320 nm and emission at 620 and 665 nm). The TR-FRET ratio was determined as the fluorescence of the acceptor (665 nm) over the fluorescence of the donor (620 nm). The Lance Ultra cAMP assay was performed according to the manufacturer's recommendations. Briefly, 1,000 cells per well (384-well shallow well plate) were treated with increasing concentrations of NT, NT(8-13), or NN for 30 min in the presence or absence of 1 µM forskolin. cAMP-tracer and anti-cAMP-Cryptate were added for at least 1 h. The plates were read on a GENios Pro plate reader with HTRF filters, and the TR-FRET ratio was calculated as described for the IP-One assay.

*2.6 AlphaScreen assay*

ERK1/2 phosphorylation in the CHO-hNTS1 cells was monitored with Perkin Elmer's AlphaScreen SureFire assays as previously described (Koole et al., 2010). Cells were seeded in 96-well plates. Twenty-four hours later, the cells were serum-starved overnight. For concentration-response curves, the cells were stimulated for 10 min with increasing concentrations of NT, NT(8-13), or NN. For time-course curves, the cells were incubated with inhibitors as mentioned in the figure legend and stimulated for 0, 5, 10, 15, 30, 45, and 60 min with 1 µM NT. The cells were then lysed with 20 µL of 5X lysis buffer, incubated at room temperature for 10 min on a plate shaker, and frozen overnight at -20°C. A total of 5 µL of the lysate was used for analysis. The readings were performed on a Perkin Elmer EnSpire 2300 multilabel reader.

*2.7 Whole-cell integrated response quantification*

The measurements of the cell monolayer resistivity were performed in cysteine-stabilized, Poly-L-Lysine-coated 96-well plates (96w20idf from Applied Biophysics, Troy, NY, USA) as recommended by the manufacturer. A total of 30,000 CHO-hNTS1 cells were plated into each well and allowed to grow for 24 h in serum-containing media before being serum-starved for 16 to 18 h prior to the experiment. On the day of the experiment, the cells were washed with PBS, and 90 µL of Leibovitz's L-15 modification medium (L15; Wisent, St-Bruno, QC, Canada) was then added per well. The biophysical parameters (electrical resistance at a 4000 Hz single frequency) of the cell monolayer were recorded with an ECIS Zθ linked to a 96-well array station (Applied Biophysics, Troy, NY, USA) for 60 min prior to stimulation with NT, NT(8-13) or NN at 1, 10, 100 or 1000 nM or in the presence of the NTS1 receptor antagonist SR48692 at 10 µM.

The whole-cell integrated behaviors were then monitored for a total of 180 min following stimulation.

Two days before surface plasmon resonance (SPR) detection, the cells were seeded into each well containing the SPR substrate (the preparation of the SPR substrate was described by Cuerrier et al., (2008) at a density of 100,000 cells/cm² and serum-starved the following day. Before the SPR signal was measured, the cells were washed twice with L15 medium and transferred to the SPR apparatus. Ninety percent confluency, as evaluated by contrast microscopy, is necessary in order to perform the experiment. The SPR analysis was performed at 37°C on a custom-built surface plasmon resonance apparatus as previously described (Chabot et al., 2009; Cuerrier et al., 2008).

## 2.8 Data analysis

The $EC_{50}$ values of the $BRET^2$, IP-One, Lance Ultra cAMP, AlphaScreen, and whole-cell integrated response assays were determined as the concentration of ligand showing 50% of activation. The data were calculated using concentration-response three parameters non-linear regression of GraphPad Prism 6 (La Jolla, CA, USA). The experimental numbers are described in each figure legend.

The half-times for β-arrestin recruitment were obtained using the one-phase association exponential regression in GraphPad Prism 6 and are represented as the half-time ± S.E.M. of 3 independent experiments, each done in triplicate.

Electrical resistivity traces are the averaged normalized response values ± S.E.M. of at least 3 separate experiments. The normalized response was calculated as the resistivity of the cell monolayer divided by the average resistivity during the 60 min baseline. The whole-cell integrated concentration-response curve was generated using the maximum value of the normalized response for each ligand concentration.

SPR traces are the averaged normalized reflectance values ± S.E.M. of at least 6 separate experiments. The normalized reflectance was calculated as the reflectance value at a specified time minus the reflectance value at the beginning of the experiment.

Statistical analyses were performed using GraphPad Prism 6 and are described in the figure legend when applicable. A value was considered statistically significant when $P < 0.05$.

## 3. Results

### 3.1 Agonist-dependent hNTS1 activation of $G\alpha_q$, $G\alpha_i$, $G\alpha_o$, and $G\alpha_{13}$

To elucidate the signaling signature of hNTS1 following its binding with NT, NT(8-13), or NN (sequences provided in **Table 1**), we initially assessed the ability of the receptor to couple to various G proteins using BRET biosensors in CHO-K1 cells stably expressing the receptor. We first determined the ability of hNTS1 to activate $G\alpha_q$ by measuring the dissociation of $G\alpha_q$-RlucII and GFP10-$G\gamma_1$. NT, NT(8-13), and NN promoted a decrease in the BRET$^2$ ratio, confirming the activation of $G\alpha_q$ by stimulated hNTS1 receptors (**Fig. 1A**). The activation of $G\alpha_q$ by NT, NT(8-13), and NN was found to be concentration dependent, and the EC$_{50}$ values are shown in **Table 2**. These results are in agreement with the potencies previously observed in the NTS1/$G\alpha_q$-related second messenger assays (Choi et al., 1999; Kitabgi et al., 1986). Since several studies have shown that NT-induced arachidonic acid production was dependent on the PTX-sensitive $G\alpha_{i/o}$ pathway (Gailly et al., 2000; Najimi et al., 2002), we further investigated whether NTS1 promoted the engagement of $G\alpha_{i1}$ and $G\alpha_{oA}$ in a similar experimental paradigm as the one used for $G\alpha_q$ (**Fig. 1B, C; Table 2**). Our results show that all three ligands activated both $G\alpha_{i1}$ and $G\alpha_{oA}$. To further study the signaling signature of hNTS1, we also assessed the ability of these three ligands to promote the activation of $G\alpha_{13}$, which is known to modulate small GTPase RhoA activity (Offermanns, 2003). By measuring the dissociation between $G\alpha_{13}$-RlucII and GFP10-$G\gamma_1$, we found that NT, NT(8-13), and NN promoted the activation of $G\alpha_{13}$ (**Fig. 1D, Table 2**). The concentration-response curves revealed dose-dependent activation of $G\alpha_{13}$ by NT, NT(8-13), and NN. Altogether, these results reveal for the first time the ability of hNTS1 to directly activate the $G\alpha_q$-, $G\alpha_{i1}$-, $G\alpha_{oA}$-, and $G\alpha_{13}$-mediated signaling pathways.

### *3.2 Downstream G protein signaling reveals functional coupling with G$\alpha_q$, G$\alpha_s$, and G$\alpha_i$*

To corroborate the results observed with the BRET-based assays, we monitored downstream signaling events by quantifying the production of second messengers generated by hNTS1 activation. For the G$\alpha_q$ pathway, we tested the ability of NT, NT(8-13), and NN to stimulate the inositol phosphate (IP) cascade by measuring IP1 accumulation over 30 min (**Fig. 2A**). Interestingly, NT(8-13) was more potent than NT and NN at activating the IP signaling pathway (**Fig. 2A; Table 2**). Moreover, the use of UBOQIC, a G$\alpha_q$ blocker (Inamdar et al., 2015) completely abolished the NT-induced increase in inositol phosphate formation (**Fig. 2B**), thus reinforcing the point that NT-induced calcium release is mediated solely through G$\alpha_q$ activation. Since NTS1 was found to potentiate the adenylyl cyclase-dependent production of cAMP through G$\alpha_s$ activation (Skrzydelski et al., 2003; Yamada et al., 1993), we evaluated the ability of these three NTS1 ligands to produce cAMP over a 30-min stimulation period (**Fig. 2C**). All three ligands elevated the cAMP levels, but no differences were observed for the three compounds with respect to their EC$_{50}$ values (**Table 2**). The production of cAMP is usually related to the activation of G$\alpha_s$, while G$\alpha_{i/o}$ is known to inhibit cAMP production by adenylyl cyclase. However, there is also evidence that NTS1 may trigger a G$\alpha_s$-independent increase in cAMP levels through G$\alpha_{q/11}$-mediated activation of calcium-dependent adenylyl cyclase (Carraway and Mitra, 1998). In our model, we found that the G$\alpha_q$ blocker UBOQIC did not inhibit NT-induced cAMP production, therefore reinforcing the role played by G$\alpha_s$ in cAMP accumulation (**Fig. 2D**).

To further understand the role of G$\alpha_{i/o}$ in the cell's cAMP levels, we next examined the effect of blocking G$\alpha_{i/o}$ function on the NT-induced cAMP accumulation by treating the cells with PTX. PTX is a toxin known to interfere on the G$\alpha_{i/o}$ pathway by an ADP-ribosylation of the G$\alpha$ subunit resulting in a loss of the ability of this subunit to dissociate from the Gβγ dimer, thus, uncoupling the receptor form the $G_{i/o}$ proteins(Mangmool and Kurose, 2011). Strikingly, NT was found to produce higher levels of cAMP in the presence of PTX (**Fig. 2D**). These results thus suggest that hNTS1 is functionally coupled to $G\alpha_{i/o}$ but that the $G\alpha_s$-induced cAMP production observed following NT treatment counteracts the PTX-sensitive decrease in cAMP formation.

*3.3 hNTS1 activation leads to the recruitment of β-arrestins 1 and 2*

Given the predominant role played by β-arrestins in GPCR desensitization and internalization as well as G protein-independent signaling (Pierce and Lefkowitz, 2001; Premont and Gainetdinov, 2007), we further monitored the ability of hNTS1 to recruit β-arrestins 1 and 2 following receptor activation. Indeed, it had been previously reported that hNTS1 recruits β-arrestins after NT stimulation (Oakley et al., 2001; 2000); however, both the β-arrestin recruitment kinetics and potencies were never quantified. We thus used a BRET-based assay to measure the recruitment of β-arrestins 1 and 2 to the activated receptor in the CHO-K1 cells expressing hNTS1. We first performed time-course experiments to assess the dynamics of β-arrestin recruitment to hNTS1 (**Fig. 3A, B**). Our results reveal that NT, NT(8-13) and NN promoted rapid $BRET^2$ increases between hNTS1-GFP10 and RlucII-β-arrestin 1 or 2, which indicates β-arrestin recruitment to the hNTS1 receptor. For both β-arrestins, the time required for NT, NT(8-13), and NN to reach a maximal signal was equivalent, and this maximal response was sustained over time. The recruitment of β-arrestin 1 following NT, NT(8-13), or NN stimulation was completed with half times of 61.6 ± 5.9 s, 61.5 ± 7.9 s, and 64.9 ± 8.8 s, respectively. β-arrestin 2 recruitment induced through NTS1 activation by NT, NT(8-13), or NN was accomplished in half times of 54.8 ± 8.2 s, 76.5 ± 11.9 s, and 104 ± 20 s, respectively. The recruitment of β-arrestin 1 induced by NTS1 activation was found to be concentration-dependent (**Fig. 3C**), and the $EC_{50}$ values are summarized in **Table 2**. NT and NT(8-13) were also found to induce β-arrestin 2 recruitment with a higher potency than NN (**Fig. 3D**; **Table 2**). Given the role of β-arrestins in receptor endocytosis, our results on β-arrestin 1 and 2 recruitment suggest that NT, NT(8-13), and NN will lead to hNTS1 receptor endocytosis (Claing et al., 2002; Ferguson et al., 1996).

*3.4 hNTS1 activates ERK1/2 through distinct G protein families*

We next investigated whether NT, NT(8-13), or NN induced the activation of ERK1/2 upon hNTS1 stimulation in our cellular model. Therefore, CHO cells stably expressing hNTS1 were stimulated with increasing concentrations of NT, NT(8-13), or NN for 10 min prior to the addition of lysis buffer. As shown in **Fig. 4A**, NT, NT(8-13), and NN promoted ERK1/2-phosphorylation in a concentration-dependent manner, and the $EC_{50}$ values are reported in **Table 2**. Since ERK1/2 activation can occur through several intracellular mechanisms depending on receptor and cell type, we further investigated the contributions of signaling via G protein dependent and independent pathways to the activation of the ERK pathway.

To this end, we used different pharmacological inhibitors to determine which downstream signaling pathways mediate the hNTS1 stimulation-induced ERK1/2 phosphorylation. Since several studies have reported that GPCRs may mediate ERK1/2 activation through an EGFR transactivation-dependent mechanism (Overland and Insel, 2015), we first assessed whether the cell-permeable potent selective EGFR inhibitor PD168393 would decrease ERK1/2 phosphorylation. Our results revealed that EGFR inhibition did not change the ERK1/2 phosphorylation profile when compared to untreated cells, thereby suggesting that EGFR transactivation is not involved in ERK1/2 activation in our cellular model (**Fig. 4B**). Recent evidence also suggests that β-arrestins can act as signal transduction scaffolds for ERK1/2 activation following agonist-stimulated receptor internalization (Reiter and Lefkowitz, 2006). We thus examined if the use of dynasore, a small molecule GTPase inhibitor that targets dynamin and blocks endocytosis, could lead to a decrease in ERK1/2 phosphorylation (**Fig. 4B**). We did not observe significant changes in NT-induced ERK1/2 activation in the presence of dynasore, suggesting that NTS1-induced ERK1/2 phosphorylation is not dependent on β-arrestin internalization.

We next investigated the contributions of the $G\alpha_{13}$, $G\alpha_q$ and $G\alpha_{i/o}$, proteins to the hNTS1-mediated activation of the ERK1/2 pathway. The involvement of $G\alpha_{13}$ in ERK1/2 phosphorylation was determined by exposing the CHO-K1 cells expressing hNTS1 to a 30 min pretreatment with 10 µM Y27632, a cell-permeable and highly selective inhibitor of $G\alpha_{13}$-induced Rho-associated kinases (ROCK) activation (Narumiya et al., 2000). No significant differences in ERK1/2 phosphorylation were detected following Y27632 pretreatment (**Fig. 4B**). We next used the UBOQIC blocker and the potent PKC inhibitor Gö6983 to evaluate the roles of $G\alpha_q$ protein and the subsequent downstream phospholipase Cβ-DAG-PKC pathway, respectively, in ERK1/2 activation. Both inhibitors were found to reduce the long-lasting effect of NT on ERK1/2 phosphorylation, suggesting that hNTS1-induced ERK1/2 activation involves the sequential activation of $G\alpha_q$ and PKC (**Fig. 4C**).

Finally, GPCRs are capable of inducing ERK1/2 activation via PTX-sensitive G proteins with Src kinase and Gβ/γ-dependent mechanisms (Rozengurt, 2007). The contribution of $G\alpha_{i/o}$ protein to hNTS1-mediated ERK1/2 activation was therefore determined following an overnight pretreatment with PTX as a $G\alpha_{i/o}$ blocking agent. Despite the demonstration that hNTS1 triggered ERK1/2 activation through $G\alpha_q$ (**Fig. 4C**), a complete loss of ERK1/2 phosphorylation was observed in PTX-treated cells (**Fig. 4D**), thus highlighting the key role for $G\alpha_{i/o}$ in NT-induced ERK1/2 activation. Since PTX pretreatment induces ADP-ribosylation of $G\alpha_{i/o}$ proteins and locks them into their inactive heterotrimeric conformation, we then evaluated whether $G\alpha_{i/o}$-mediated ERK1/2 activation was induced by a $G\beta/\gamma$-dependent mechanism by treating cells with gallein, a $G\beta/\gamma$ signaling inhibitor (Bonacci et al., 2006; Smrcka, 2013). No differences in the ERK1/2 phosphorylation rate between gallein-treated and untreated cells were recorded, suggesting that NT-induced ERK1/2 activation is mediated by $G\alpha_{i/o}$ activity and not by $G\beta/\gamma$-mediated signaling (**Fig. 4D**). We finally tested whether $G\alpha_{i/o}$-induced ERK1/2 activation was mediated through downstream activation of the proto-oncogene tyrosine-protein kinase c-Src. This non-receptor tyrosine kinase can be activated by $G\alpha_{i/o}$ and leads to ERK1/2 activation through direct phosphorylation of Ras/Raf or by indirect activation that involves EGFR transactivation (D. K. Luttrell and L. M. Luttrell, 2004; New and Wong, 2007; Piiper and Zeuzem, 2004). As shown in **Fig. 4D**, the inhibition of c-Src by a 30-min pretreatment with 1 µM PD166285, a non-selective c-Src and receptor tyrosine kinase inhibitor (Panek et al., 1997), resulted in a loss of more than 75% of the ERK1/2 activation as well as in the appearance of a delay in the ERK1/2 phosphorylation rate. Since we had previously demonstrated that ERK1/2 phosphorylation was independent of EGFR activation, the results obtained with PD166285 suggest that in our cellular model, hNTS1-mediated ERK1/2 activation may result from direct c-Src activation by $G\alpha_{i/o}$.

*3.5 Whole-cell integrated responses following hNTS1 activation*

Because NTS1 activates several signaling pathways, we next used impedance-based cell monitoring approach to monitor and quantify the whole-cell integrated response resulting from the activation of hNTS1 with NT, NT (8-13) or NN. Impedance monitoring is a noninvasive label-free technique that allows the real-time measurement of dynamic mass redistribution (DMR) in living cells grown on gold-plated electrodes and exposed to different ligands. As shown in **Fig. 5 A-F**, NT, NT(8-13), and NN produced a similar response profile over time, for each ligand concentration tested (i.e. 1, 10, 100, and 1000 nM), in which at least two distinct phases and one transient plateau could be observed. Immediately after the stimulation of CHO-hNTS1 cells, a rapid increase in the response signal was observed until it reached a transient maximum value followed by a plateau, which is only present for stimulations at 100 and 1000 nM. This maximum was then pursued by a slow decrease of the response until a return to the baseline at 180 min after stimulation. NT induced a more sustained response than NT(8-13) and NN at high concentrations (100 and 1000 nM) but the global shape of the curve was not affected (**Fig. 5 C-D**). As a control, L15 media was applied to CHO-hNTS1 cells, which did not produce any variations in the normalized response aside from the injection peak (**Fig. 5E**). As expected, the response of CHO-hNTS1 cells stimulated with 10 µM of the NTS1 receptor antagonist SR48692 did not show any changes in the normalized response (**Fig. 5G**). A whole-cell integrated concentration-response curve was generated using the maximum value of the normalized response for each ligand and is shown in **Fig. 5G**. The cellular response induced by either NT, NT(8-13) or NN was comparable, even at non-saturating ligand concentrations and no significant differences were observed between the $EC_{50}$ values of the whole-cell integrated response triggered by these three ligands (**Table 2**). To corroborate the results obtained using the impedance-based cell response, we next used cell-based surface plasmon resonance (SPR) to monitor the whole-cell integrated response resulting from the activation of hNTS1 with 1 µM NT, NT (8-13) or NN. Like the impedance-based method, SPR is a noninvasive label-free technique that allows to quantify DMR in living cells grown on SPR substrate; however, this technique is based on an optical measurement instead of an electrical measurement. As shown in **Fig. 6 A-F**, NT, NT(8-13), and NN produced an SPR response profile in which at least three distinct phases and two transient plateaus could be observed. As a control, NT was applied to CHO-K1 cells (Mock cells) (**Fig. 6A**) or L15 media was applied to CHO-hNTS1 cells (**Fig. 6B**), which did not produce any variations in the SPR signal aside from the injection peak. Prior to stimulation, the SPR signal was stable, indicating a steady-state level of cellular activity. Immediately after the stimulation of CHO-hNTS1 cells, a rapid decline in the SPR signal was observed until it reached a transient minimum value (-6.8 ± 1.0 reflectance variation unit (RVU), -4.2 ± 1.0 RVU and -5.0 ± 1.2 RVU for NT, NT(8-13) and NN, respectively), which was identified as phase I. We next observed an increase in the SPR signal (phase II), which could be quantified by calculating the difference between the minimum and the first observed plateau when cell response progressed toward a transient stable state after stimulation (22.0 ± 1.2 RVU for NT, 20.2 ± 2.1 RVU for NT(8-13) and 21.0 ± 0.8 RVU for NN). Following this first plateau, another increase was observed and was identified as phase III. This third phase was quantified by calculating the difference between the value at the first plateau and the value at the second plateau after stimulation (13.7 ± 2.8 RVU, 11.7 ± 3.1 RVU and 13.8 ± 1.3 RVU for NT, NT(8-13) and NN, respectively). No significant differences were observed between all phases of the SPR response (**Fig. 6F**).

DMR quantification experiments (changes in cell monolayer impedance and SPR) thus revealed similar cellular behaviors in response to NT, NT(8-13) or NN stimulation, suggesting that the integrated signatures triggered by NTS1 activation lead to similar cellular responses.

## 4. Discussion

In the present study, we report the signaling signature of the hNTS1 receptor in CHO-K1 cells following its activation by NT, NN, and NT(8-13) using BRET-based and label-free whole-cell DMR sensing assays (changes in cell monolayer impedance and SPR). The characterization of the hNTS1-mediated cellular signaling network revealed that NT, NT(8-13), and NN activate the $G\alpha_q$-, $G\alpha_{i/o}$-, $G\alpha_{13}$-, and $G\alpha_s$-protein signaling pathways with similar efficacies and potencies. Furthermore, all ligands stimulated the production of inositol phosphate, modulated cAMP accumulation as well as activated ERK1/2 in our cellular model (See **Fig. 7** for a comprehensive scheme of the signaling pathways activated by hNTS1). Nevertheless, we noticed that NN was less potent than NT and NT(8-13) for inducing the recruitment of β-arrestins, whereas NT(8-13) was more potent than NT and NN in activating the IP signaling pathway. Despite these distinct signaling profiles toward different effector systems, the integrated signatures revealed by two different label-free whole-cell DMR sensing assays (i.e. changes in cell monolayer impedance and SPR; **Fig. 5** and **Fig. 6**) led to identical cellular responses by the three ligands.

Interaction between hNTS1 and G proteins have been mainly based on the activation of second messenger signaling pathways and not on a direct assessment of the physical interactions between the receptor and the heterotrimeric G proteins. Our results reveal that hNTS1 stimulation promoted the engagement of different G protein families at the G protein level, which was previously demonstrated by their coupling to various downstream signaling pathways and by the use of signal transduction inhibitors (Pelaprat, 2006; Vincent et al., 1999; Z. Wu et al., 2012). We found that hNTS1 stimulation induced $G\alpha_q$ activation and the subsequent production of inositol phosphate. The $G\alpha_q$-dependent pathway was confirmed by the application of the $G\alpha_q$ inhibitor UBOQIC.

These results are consistent with the previous findings in other cell types, which show that NTS1, via its coupling to the $G\alpha_q$ subunit, mediates the activation of phospholipase C and the production of IP and diacylglycerol, leading to the mobilization of intracellular calcium and activation of PKC in both neuronal and non-neuronal cell types (Amar et al., 1987; Choi et al., 1999; Najimi et al., 2002; Oury-Donat et al., 1995; Schaeffer et al., 1995; Skrzydelski et al., 2003; Snider et al., 1986; Wang and T. Wu, 1996).

NTS1-induced $G\alpha_{i/o}$ activation was previously suggested by experimental studies that showed the PTX-dependent sensitivity of [$^{35}$S]-GTPγS binding following NT activation (Gailly et al., 2000) and by inhibiting cAMP production in N1E115 cells (Bozou et al., 1986; Oury-Donat et al., 1995). Reports have also shown that NT stimulated cAMP production in heterologous cells expressing human or rat NTS1 as well as in pancreatic cancer cells endogenously expressing hNTS1, indicating that NTS1 also couples to $G\alpha_s$ (Ishizuka et al., 1993; Richard et al., 2001; Sarret et al., 2010; Skrzydelski et al., 2003; Yamada et al., 1993). Here, our results provide a new insight into the discrepancy that demonstrated both $G\alpha_{i/o}$- and $G\alpha_s$-NTS1 coupling (**Fig. 2**). Indeed, we demonstrate that NT-induced cAMP production was increased in the presence of PTX. Since $G\alpha_s$ and $G\alpha_{i/o}$ regulate the same intracellular effector, these results suggest that in this host cell system, hNTS1 is preferentially coupled to adenylyl cyclase through $G\alpha_s$ rather than $G\alpha_{i/o}$, consequently leading to cAMP production. It has also been suggested that NTS1 causes an increase in the cellular level of cAMP through the $G\alpha_{q/11}$-mediated activation of calcium-dependent adenylyl cyclase in PC3 cancer cells (Carraway and Mitra, 1998). Here, we found that the $G\alpha_q$ blocker UBOQIC did not affect NT-induced cAMP accumulation. These data thus reinforce the central role played by $G\alpha_s$ in cAMP production.

We also demonstrate the ability of NTS1 to engage G$\alpha_{13}$ (**Fig. 1**). This finding is in line with reports showing that NT modulates the activity of small Rho GTPases, such as Cdc42 and Rac1, downstream of G$\alpha_{13}$ in U373 glioblastoma cells (Servotte et al., 2006). These results further suggest that NTS1 activation could mediate cytoskeletal rearrangements and lead to the emergence of stress fibers and cell migration (Moura et al., 2013).

β-Arrestins bind agonist-activated receptors and play a central role in mediating GPCR desensitization and internalization (Ferguson, 2001; Pierce and Lefkowitz, 2001). We therefore investigated the effects of hNTS1 stimulation on G protein-independent signaling by monitoring the recruitment β-arrestins 1 and 2 to the receptor (**Fig. 3**). Our results of hNTS1-triggered β-arrestin recruitment indicate rapid binding of both β-arrestins to the receptor when subnanomolar levels of ligands are used. The cellular trafficking of β-arrestins is differentially regulated via their association to distinct GPCR subtypes; class A receptors interact with β-arrestin 2 with higher affinities than β-arrestin 1, while class B receptors bind both β-arrestins with similarly high affinities (Drake et al., 2006). This type of β-arrestin-receptor interaction appears to determine receptor fate; GPCRs that are tightly bound to β-arrestins (class B) are more frequently targeted to lysosomal compartments where they undergo proteolytic degradation, whereas receptors dissociating rapidly from β-arrestins (class A) are recycled back to the plasma membrane (Drake et al., 2006). Our results support the classification of hNTS1 as a class B GPCR (Oakley et al., 2001; 2000; Zhang, 1999) and are consistent with previous observations demonstrating that hNTS1 receptors are trafficked to lysosomal compartments for degradation (Vandenbulcke et al., 2000). Our results also reveal that both NT and NT(8-13) were more potent in inducing the recruitment of β-arrestins compared to NN. These differences might be driven by the presence of GRK kinases, which have been shown to influence GPCR signaling and trafficking (Liggett, 2011). Accordingly,

Inagaki et al. (2015) recently reported that GRK2 and GRK5 differentially phosphorylate the NTS1 receptor. Phosphorylation of NTS1 by GRK2 is agonist-independent and occurs only on the C-terminal Ser residues, whereas phosphorylation of NTS1 by GRK5 is agonist-dependent and requires the phosphorylation of the Ser and Thr residues located in intracellular loop 3 and the C-terminus. Since these GPCR kinases seem to be endogenously expressed in CHO-K1 cells (Osorio-Espinoza et al., 2014), we can therefore speculate that they could affect the responses induced by the different NT agonists and lead to different profiles of β-arrestin recruitment.

NTS1 was found to be overexpressed in several cancer cell lines (Mustain et al., 2011), such as colonic adenocarcinomas (Evers, 2006; Müller et al., 2011; Yoshinaga et al., 1992; Zhao et al., 2007), breast (Dupouy et al., 2009; Souazé et al., 2006), and lung cancers (Allen et al., 1988; Bunn et al., 1992; Davis et al., 1991). Thus, we were also interested in investigating whether NTS1 signaling by NT, NT(8-13), and NN led to activation of the MAPK proliferative pathway. Since many signaling pathways, including G protein-dependent and G protein-independent pathways, lead to ERK1/2 phosphorylation, the pharmacological inhibitors of upstream signaling molecules were used to decipher the contribution of the various effectors involved (**Fig. 4**). In contrast to previous studies demonstrating that the EGFR inhibitor gefitinib inhibited NT-induced ERK1/2 phosphorylation in colonic HT-29 and prostatic PC3 cancer cell lines (Hassan et al., 2004; Müller et al., 2011), we did not observe that EGFR transactivation was required for ERK1/2 activation. These discrepancies may be related to the fact that the cancer cell lines express the third member of the NT receptor family NTS3 (Morinville et al., 2004; Sarret et al., 2001). Indeed, as observed by (Martin et al., 2002), co-expression of NTS3 with NTS1 significantly affects NT-dependent intracellular signaling, such as ERK1/2 phosphorylation. Our results also reveal that ERK1/2 activation does not involve G$\alpha_{13}$ recruitment since the highly selective ROCK inhibitor Y27632 did not induce any changes in ERK1/2 phosphorylation. Several lines of evidence suggest that in addition to their role in desensitization/internalization, β-arrestins act as signal transduction scaffolds for ERK1/2 activation following agonist-stimulated receptor internalization (Pierce and Lefkowitz, 2001). Since hNTS1 internalizes via clathrin-coated pits and through a dynamin-dependent mechanism (Savdie et al., 2006; Vandenbulcke et al., 2000; Vincent et al., 1998), we further examined if dynasore that interferes with the GTPase activity of dynamin and blocks receptor endocytosis exerted an inhibitory action on ERK1/2 activation. No significant changes in NT-induced ERK1/2 phosphorylation were observed in the presence of dynasore, thus suggesting that hNTS1-mediated ERK1/2 activation occurs independent of hNTS1 internalization and of β-arrestin signaling in this cell type.

Depending on the receptor and cell type, ERK1/2 activation may be mediated by PTX-sensitive or PTX-insensitive G proteins. We demonstrate here that $G\alpha_q$ inactivation by the UBOQIC blocker or inhibition of the downstream PKC by Gö6983 partially blocked ERK1/2 phosphorylation. These results thus support the involvement of the PTX-insensitive $G\alpha_q$-PLCβ-DAG-PKC pathway in hNTS1-induced ERK1/2 activation. This finding is in agreement with previous findings showing that NT stimulates ERK1/2 phosphorylation through a PKC-Raf-1-dependent signaling pathway (Guha et al., 2003). Our results also reveal the key role played by $G\alpha_{i/o}$ in hNTS1-mediated ERK1/2 phosphorylation. Indeed, locking $G\alpha_{i/o}$ in its inactive conformation by PTX pretreatment led to a complete loss of ERK1/2 phosphorylation. This PTX-sensitive effect of ERK1/2 activation is $G\alpha_{i/o}$ dependent since the Gβ/γ inhibitor gallein does not affect the ERK1/2 phosphorylation induced by NT. These data are consistent with other studies reporting that NT stimulates the transcription of the immediate-early gene *Krox-24*, a target of the ERK1/2 pathway, through a PTX-sensitive pathway (Poinot-Chazel et al., 1996).

The most unexpected result found here is that $G\alpha_{i/o}$ inhibition by PTX completely prevents hNTS1-induced ERK1/2 phosphorylation (**Fig. 4**). This finding is indeed surprising since we have demonstrated that ERK1/2 activation is also regulated in a $G\alpha_q$-PLCβ-DAG-PKC-dependent manner. This result may be explained by the fact that PTX, which blocks the $G\alpha_{i/o}$-mediated inhibition of adenylyl cyclase thereby promoting cAMP production, would inhibit the activation of Raf-1 (Piiper et al., 2000). Raf-1 inhibition would reduce the hNTS1-induced ERK phosphorylation through the PTX-insensitive $G\alpha_q$-PLCβ-DAG-PKC pathway (**Fig. 7**). This hypothesis is supported by our demonstration that inactivation of $G\alpha_{i/o}$ by PTX increases the intracellular cAMP level in CHO-hNTS1 cells in response to NT (**Fig. 2**). Furthermore, the absence of complete inhibition of ERK1/2 activation using the c-Src inhibitor PD166285 further reinforces the involvement of the PTX-insensitive $G\alpha_q$-PLCβ-DAG-PKC-Raf-1-dependent pathway in ERK1/2 activation. We can thus conclude that the full activation of ERK1/2 in CHO-hNTS1 cells by NT requires signaling through both a PTX-sensitive $G\alpha_{i/o}$-c-Src signaling pathway and a PLCβ-DAG-PKC-Raf-1-dependent pathway downstream of $G\alpha_q$ (**Fig. 7**).

Altogether, our results demonstrate the functional coupling of agonist-stimulated hNTS1 to the G proteins $G\alpha_q$, $G\alpha_{i1}$, $G\alpha_{oA}$, and $G\alpha_{13}$ as well as its ability to recruit β-arrestins and activate the IP1, cAMP, and ERK1/2 second messenger signaling cascades. The BRET-based and label-free whole-cell sensing tools used here to decipher the NTS1 signaling signature may also be helpful in accelerating the validation of potential NTS1 biased ligands. Biased signaling of GPCRs is of growing interest to drug discovery, particularly in the quest to identify compounds with high therapeutic potential while reducing unwanted side effects (Correll and McKittrick, 2014; Garland, 2013; Kenakin, 2012). Our study will form the basis for the future development of therapeutics targeting hNTS1 in clinical indications such as pain management.

**Acknowledgments**

We gratefully acknowledge the Canadian Institute of Health Research (CIHR), National Science and Engineering Research Council of Canada (NSERC), and the FRQ-S funded Réseau Québécois de Recherche sur le Médicament (RQRM) for their financial support. EBO was supported by a research fellowship from the Institut de Pharmacologie de Sherbrooke (IPS) and Centre d'Excellence en Neurosciences de l'Université de Sherbrooke (CNS). PS holds a Canada Research Chair in Neurophysiopharmacology of Chronic Pain. Drs M. Bouvier, T. Hebert, S.A. Laporte, G. Pineyro, J.-C. Tardif and E. Thorin (CQDM Team) are also acknowledged for providing us with the biosensors. The authors declare no conflicts of interest.

This is the accepted (postprint) version of the following article: Besserer-Offroy É, *et al.* (2017). Eur J Pharmacol. doi: 10.1016/j.ejphar.2017.03.046, which has been accepted and published in its final form at http://www.sciencedirect.com/science/article/pii/S0014299917302157Mustain, W.C., Rychahou, P.G., Evers, B.M., 2011. The role of neurotensin in physiologic and pathologic processes. Curr Opin Endocrinol Diabetes Obes 18, 75–82. doi:10.1097/MED.0b013e3283419052

Müller, K.M., Tveteraas, I.H., Aasrum, M., Ødegård, J., Dawood, M., Dajani, O., Christoffersen, T., Sandnes, D.L., 2011. Role of protein kinase C and epidermal growth factor receptor signalling in growth stimulation by neurotensin in colon carcinoma cells. BMC Cancer 11, 421. doi:10.1186/1471-2407-11-421

Najimi, M., Gailly, P., Maloteaux, J.M., Hermans, E., 2002. Distinct regions of C-terminus of the high affinity neurotensin receptor mediate the functional coupling with pertussis toxin sensitive and insensitive G-proteins. FEBS Lett. 512, 329–333.

Narumiya, S., Ishizaki, T., Ufhata, M., 2000. Use and properties of ROCK-specific inhibitor Y-27632, in: Regulators and Effectors of Small GTPases - Part D: Rho Family, Methods in Enzymology. Elsevier, pp. 273–284. doi:10.1016/S0076-6879(00)25449-9

New, D.C., Wong, Y.H., 2007. Molecular mechanisms mediating the G protein-coupled receptor regulation of cell cycle progression. J Mol Signal 2, 2. doi:10.1186/1750-2187-2-2

Oakley, R.H., Laporte, S., Holt, J.A., Barak, L.S., Caron, M.G., 2001. Molecular determinants underlying the formation of stable intracellular G protein-coupled receptor-beta-arrestin complexes after receptor endocytosis*. J. Biol. Chem. 276, 19452–19460. doi:10.1074/jbc.M101450200

Oakley, R.H., Laporte, S.A., Holt, J.A., Caron, M.G., Barak, L.S., 2000. Differential affinities of visual arrestin, beta arrestin1, and beta arrestin2 for G protein-coupled receptors delineate two major classes of receptors. J. Biol. Chem. 275, 17201–17210. doi:10.1074/jbc.M910348199

Offermanns, S., 2003. G-proteins as transducers in transmembrane signalling. Progress in© 2017. This manuscript version is made available under the CC-BY-NC-ND 4.0 license http://creativecommons.org/licenses/by-nc-nd/4.0/35

**Figures**

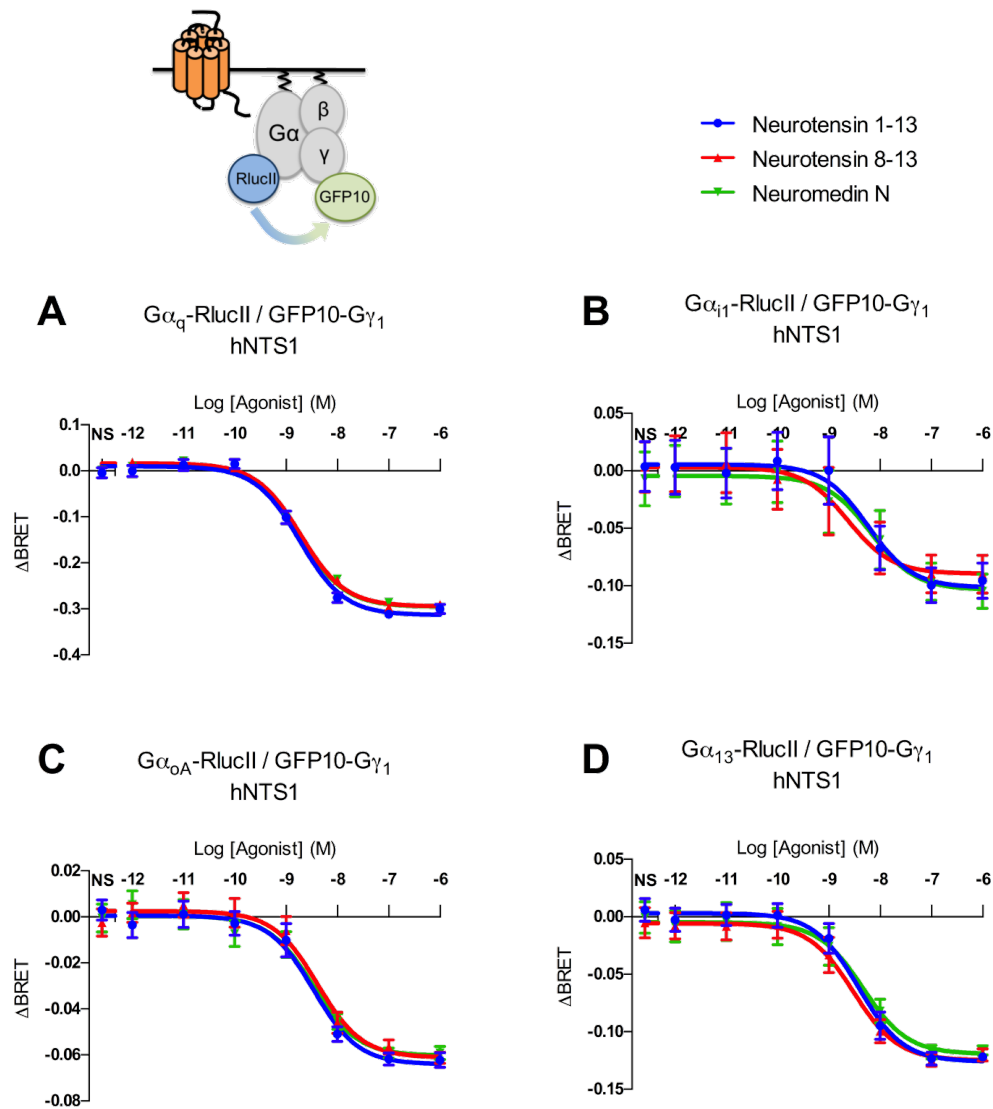

**Fig. 1. hNTS1 promotes the engagement of G proteins G$\alpha_q$, G$\alpha_{i1}$, G$\alpha_{oA}$, and G$\alpha_{13}$ in CHO-K1 cells.** Effect of increasing concentrations of NT, NT(8-13), or NN on the dissociation of G$\alpha_q$ (A), G$\alpha_{i1}$ (B), G$\alpha_{oA}$ (C), and G$\alpha_{13}$ (D). BRET was measured 10 min after the addition of the ligands. Each data set represents the mean of three independent experiments, which were each done in triplicate, and is expressed as the mean ± S.E.M..

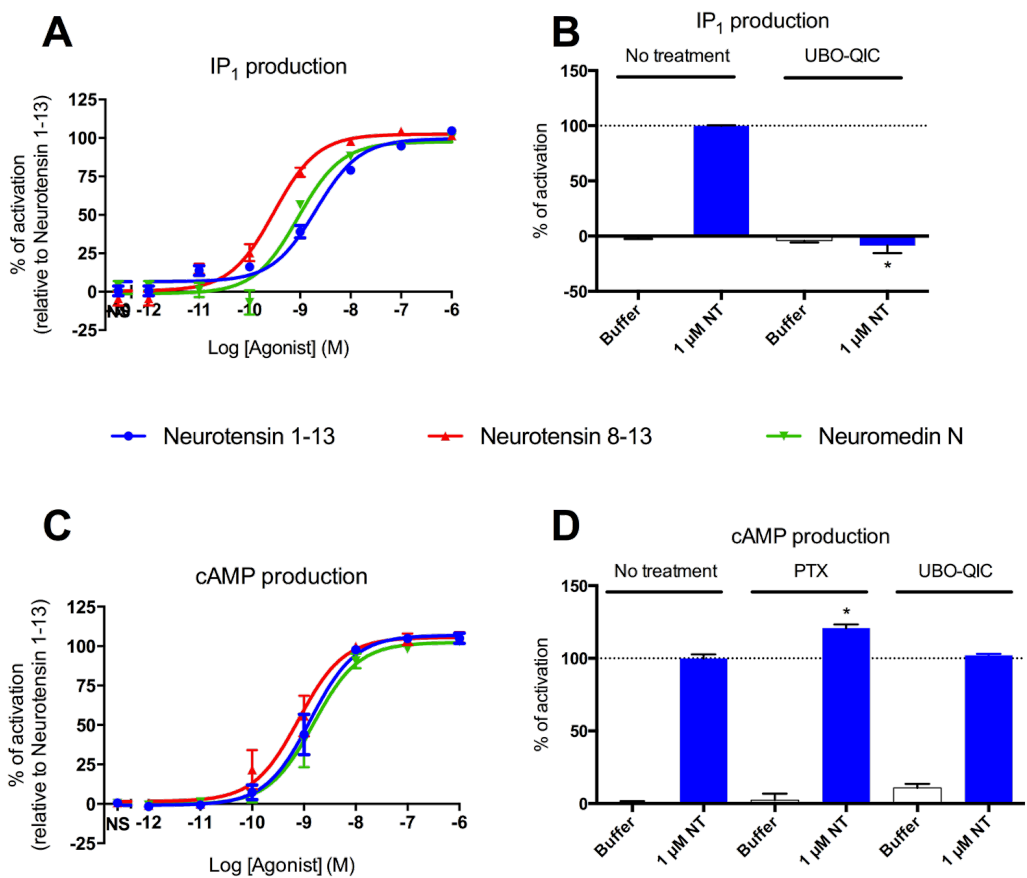

**Fig. 2. Downstream G protein signaling reveal functional coupling with $G\alpha_q$, $G\alpha_s$, and $G\alpha_{i/o}$.** (A) Effect of increasing concentrations of NT, NT(8-13), or NN on IP1 production. (B) Effect of pretreatment with the $G\alpha_q$ inhibitor UBOQIC (1 µM for 30 min) on IP1 accumulation following stimulation with 1 µM NT. (C) Effect of increasing concentrations of NT, NT(8-13), or NN on cAMP production. (D) Effect of pre-treatment with UBOQIC (1 µM for 30 min) or Pertussis toxin (PTX; 100 ng/ml, overnight) on cAMP accumulation following stimulation with 1 µM NT. * $P < 0.05$ in the Mann-Whitney test compared to the activation levels without UBOQIC or PTX treatment. Each data set represents the mean of three independent experiments, which were each performed in triplicate, and is expressed as the mean ± S.E.M..

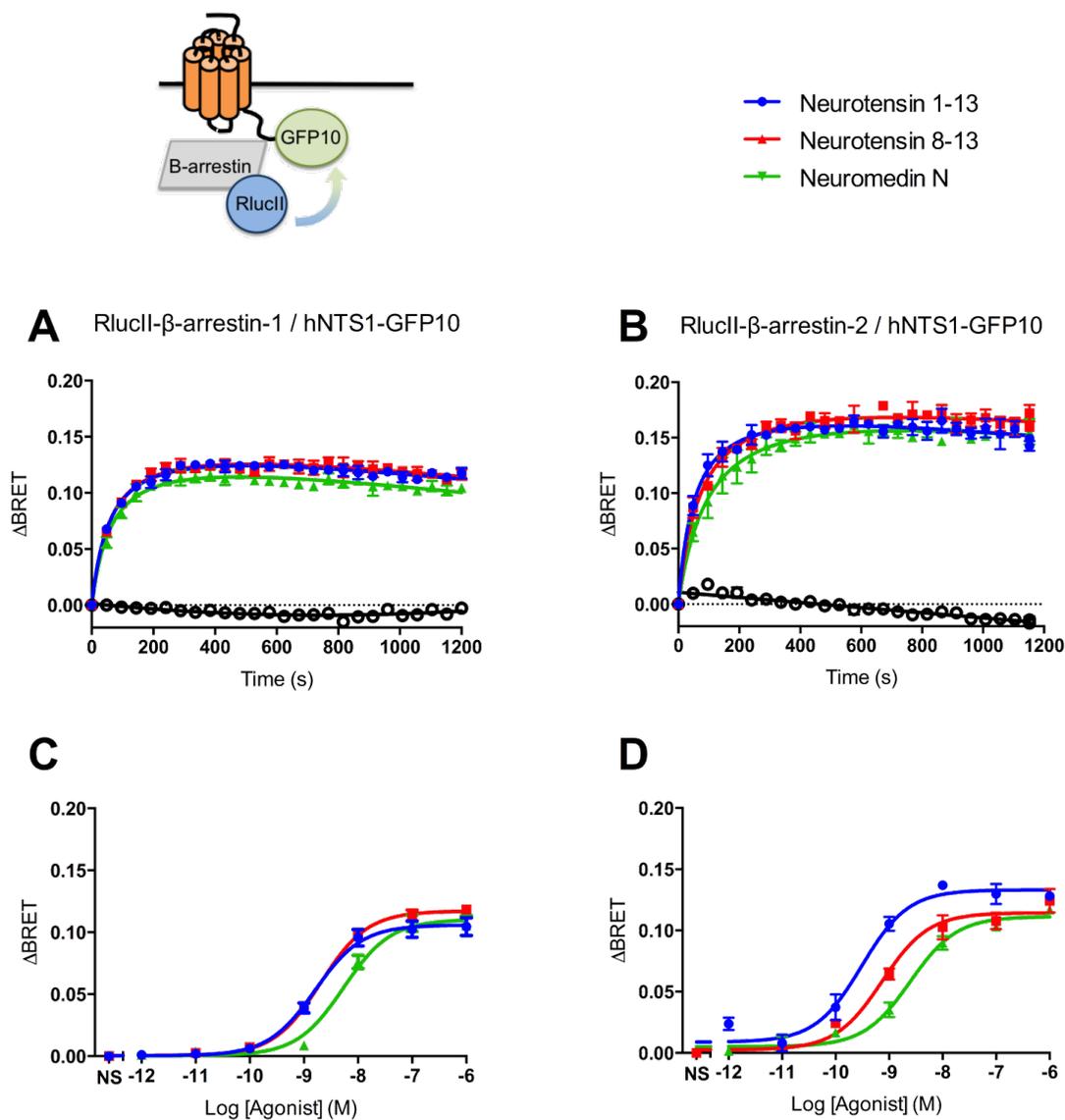

**Fig. 3. NTS1 activation leads to the recruitment of β-arrestins 1 and 2.** Kinetics of β-arrestin 1 (A) and β-arrestin 2 (B) recruitment following NTS1 activation. The black open circles represent the BRET ratio of cells stimulated only with buffer. Concentration-response curves of NT, NT(8-13), or NN on the recruitment of β-arrestin 1 (C) or β-arrestin 2 (D). Each data set represents the mean of three independent experiments, which were each done in triplicate, and expressed as the mean ± S.E.M..

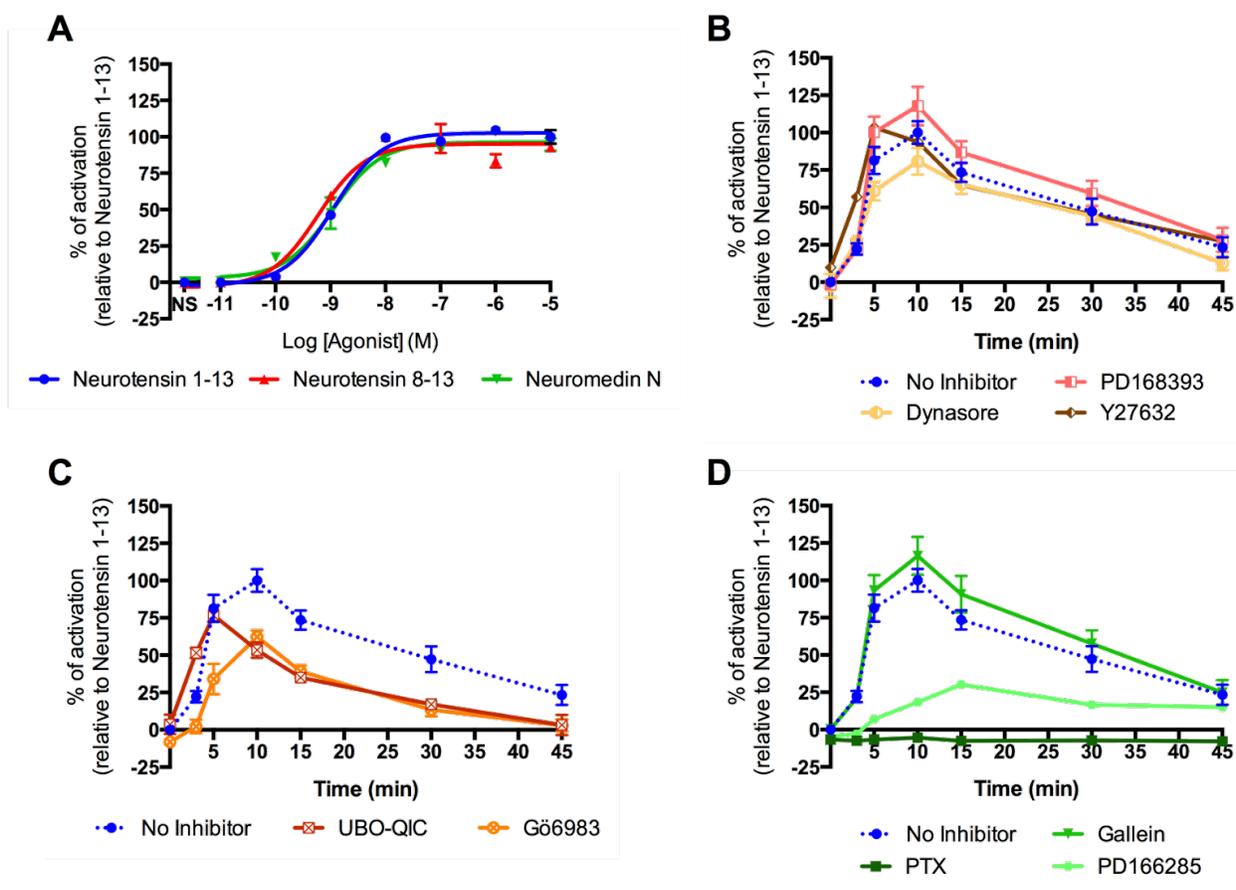

**Fig. 4. NTS1 activates ERK1/2 by a Gα$_{i/o}$-dependent mechanism.** (A) Effect of increasing concentrations of NT, NT(8-13), or NN on ERK1/2 phosphorylation. (B-D) Effects of pretreatment on ERK1/2 phosphorylation kinetics induced by 1 µM NT. (B) Pre-treatment with the EGFR inhibitor PD168393 (250 nM for 30 min), ROCK inhibitor Y27632 (10 µM for 30 min) or endocytosis inhibitor dynasore (80 µM for 30 min). (C) Pre-treatment with the Gα$_q$ blocker, UBOQIC (1 µM for 30 min) or PKC inhibitor Gö6983 (1 µM for 30 min). (D) Pre-treatment with pertussis toxin (PTX; 100 ng/ml, overnight), Gβ/γ-signaling inhibitor gallein (20 µM for 30 min) or c-Src inhibitor PD166285 (1 µM for 30 min). Each data set represents the mean of four independent experiments, which were each performed in duplicate, and is expressed as the mean ± S.E.M..

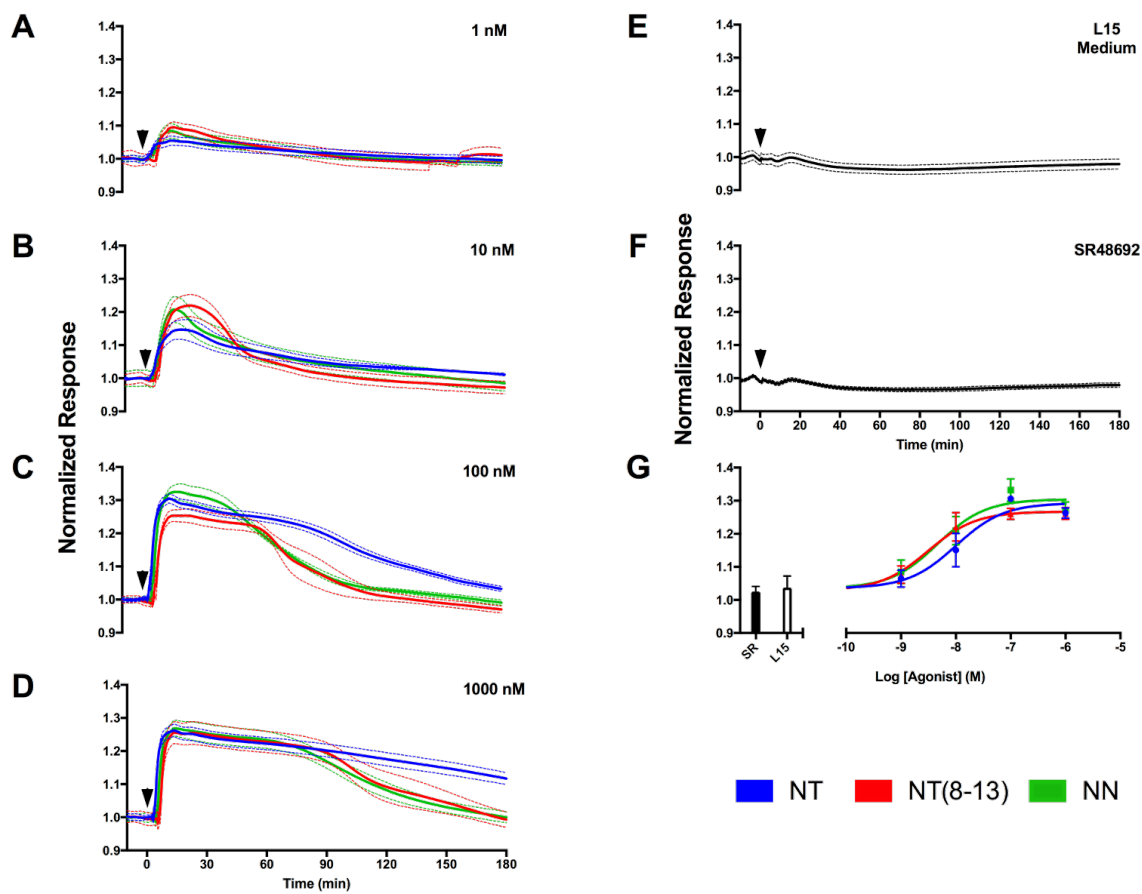

**Fig. 5. Effect of NTS1 activation on the resistivity of a cell monolayer.** (A-F) Dynamic mass redistribution kinetics assessed by cell monolayer impedance. (A-D) Normalized response of CHO-hNTS1 cells stimulated with (A) 1 nM, (B) 10 nM, (C) 100nM, and (D) 1000 nM of NT (blue line), NT(8-13) (red line), or NN (green line). (E) Normalized response of CHO-hNTS1 cells stimulated with L15 medium. (F) Normalized response of CHO-hNTS1 cells stimulated with 10 µM of the NTS1 antagonist SR48692. (G) Effect of increasing concentrations of NT, NT(8-13), or NN on the whole-cell integrated response. The arrows on each graph represent the injection time of the compound. The dotted lines represent the S.E.M. values. Each data set represents the mean of three independent experiments, which were each performed in duplicate, and is expressed as the mean ± S.E.M..

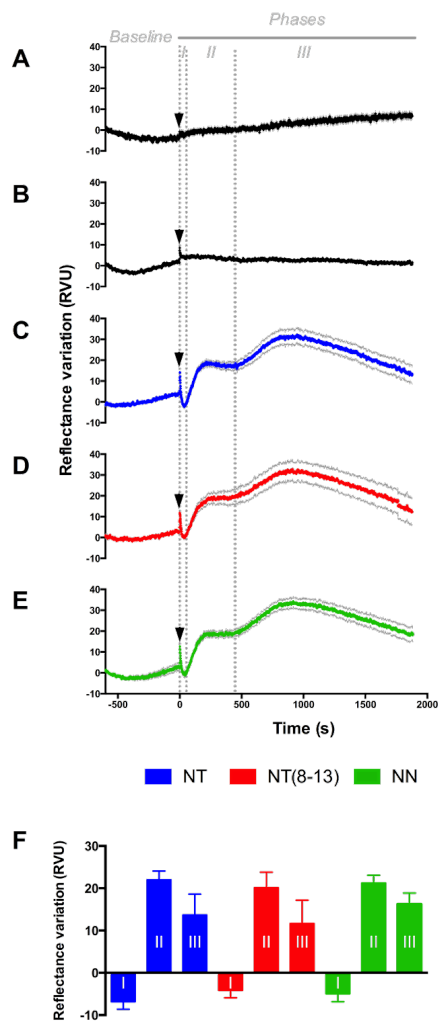

**Fig. 6. Effect of NTS1 activation on the whole-cell integrated response monitored by SPR.** (A-F) Dynamic mass redistribution kinetics assessed by SPR. (A) Reflectance variation of CHO-K1 stimulated with NT (1 µM). Reflectance variation of CHO-hNTS1 stimulated with (B) L15 medium, (C) NT (1 µM), (D) NT(8-13) (1 µM) and (E) NN (1 µM). The arrows on each graph represent the injection time of the compound. The gray dotted lines represent the S.E.M. values. (F) Reflectance variation in the RVU of phases I, II and III for each agonist. No significant differences were found between NT and NT(8-13) or NN on the Kruskal-Wallis test followed by Dunn's post hoc test. Each data set represents the mean of 6-8 independent experiments and is expressed as the mean ± S.E.M..

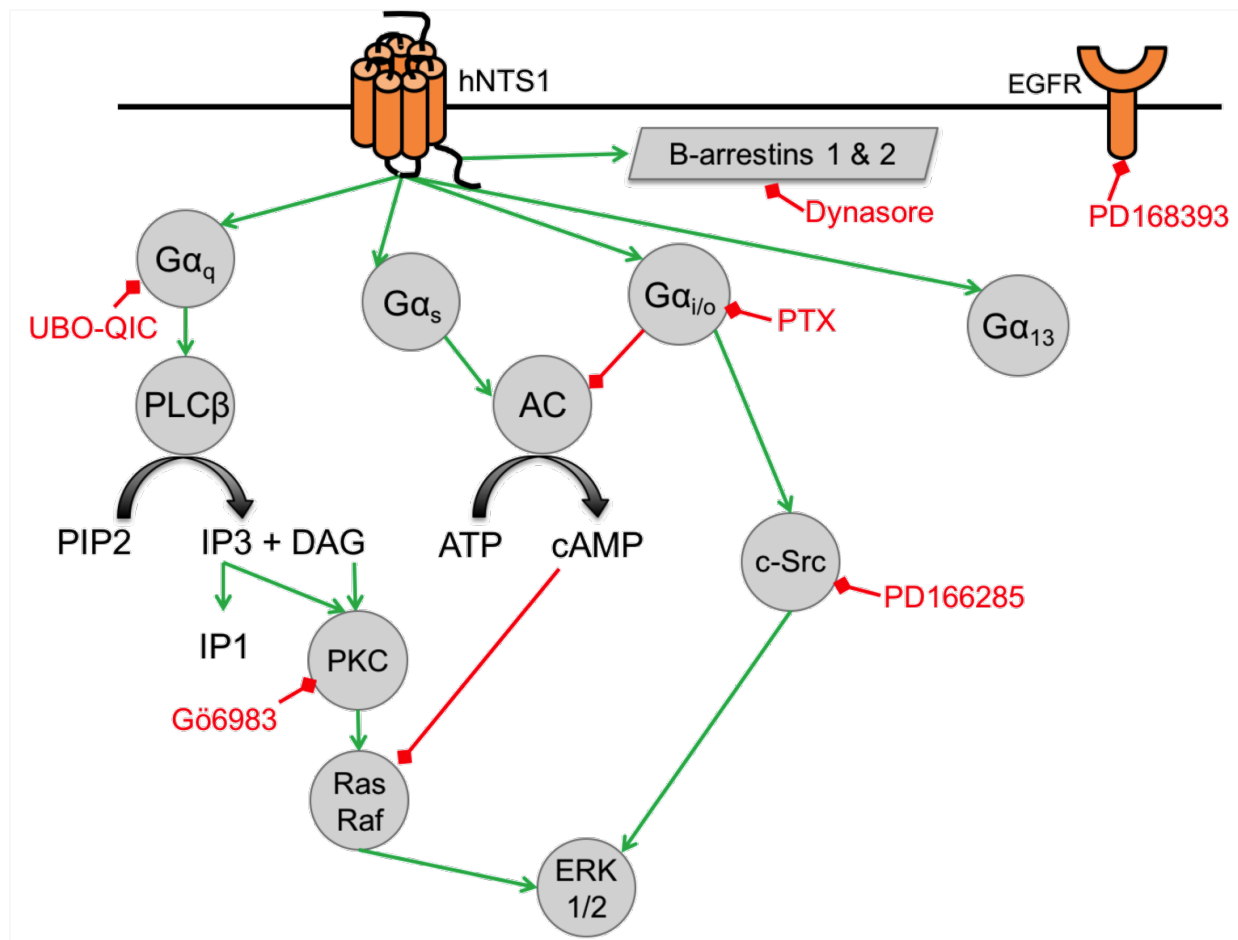

**Fig. 7. Proposed signaling signature of hNTS1 following activation by NT, NT(8-13), or NN.** Following its activation by either NT, NT(8-13), or NN, hNTS1 functionally couples to $G\alpha_q$, $G\alpha_{13}$, $G\alpha_{i/o}$, and $G\alpha_s$ and promotes the recruitment of β-arrestins 1 and 2. Furthermore, we demonstrated that in our cellular model, NTS1 was able to induce ERK1/2 phosphorylation via $G\alpha_q$- and $G\alpha_{i/o}$-dependent mechanisms. Our results showed that $G\alpha_q$-dependent signaling is involved in the late ERK1/2 response, whereas $G\alpha_{i/o}$ signaling via c-Src is responsible for the rapid increase in ERK1/2 phosphorylation. The stimulatory pathways are shown by the green arrows, and the red-squared arrows represent the inhibitory pathways and the inhibitors used in this study.

**Table 1: Peptide sequence of NT, NT(8-13) and NN**

| Compound | Sequence |
|---|---|
| Neurotensin | pGlu-Leu-Tyr-Glu-Asn-Lys-Pro-Arg-Arg-Pro-Tyr-Ile-Leu-COOH |
| Neurotensin (8-13) | H$_2$N- Arg-Arg-Pro-Tyr-Ile-Leu-COOH |
| Neuromedin N | H$_2$N- Lys-Ile-Pro-Tyr-Ile-Leu-COOH |

Table 2: $EC_{50}$ values for NT, NT(8-13) and NN on hNTS1 activated signaling pathways

| Pathway | $EC_{50}$, nM | | | | | | | | |
|---|---|---|---|---|---|---|---|---|---|
| | Neurotensin, NT | | | Neurotensin 8-13, NT(8-13) | | | Neuromedin N, NN | | |
| $G\alpha_q$ | 1.84 | ± | 0.29 | 1.92 | ± | 0.32 | 1.89 | ± | 0.34 |
| $G\alpha_{i1}$ | 5.83 | ± | 6.32 | 2.36 | ± | 3.40 | 6.64 | ± | 8.43 |
| $G\alpha_{oA}$ | 3.44 | ± | 1.26 | 4.12 | ± | 1.56 | 3.37 | ± | 1.37 |
| $G\alpha_{13}$ | 3.78 | ± | 1.37 | 3.01 | ± | 1.36 | 4.62 | ± | 2.54 |
| $IP_1$ | 0.33 | ± | 0.05 | 2.27 | ± | 0.34[b] | 0.86 | ± | 0.16[a, c] |
| cAMP | 1.32 | ± | 0.28 | 0.79 | ± | 0.22 | 1.54 | ± | 0.42 |
| β-arr-1 | 1.60 | ± | 0.29 | 2.01 | ± | 0.18 | 5.41 | ± | 1.03[b, d] |
| β-arr-2 | 0.28 | ± | 0.09 | 0.74 | ± | 0.18 | 2.01 | ± | 0.67[b, c] |
| pERK1/2 | 1.10 | ± | 0.13 | 0.60 | ± | 0.14 | 1.09 | ± | 0.25 |
| DMR | 10.1 | ± | 7.17 | 3.10 | ± | 1.81 | 4.67 | ± | 3.65 |

[a] $P<0.05$; [b] $P<0.01$ compared to NT and [c] $P<0.05$; [d] $P<0.01$ compared to NT(8-13); Kruskal-Wallis followed by a Dunn's post hoc test.